\documentclass[oneside,english,12pt]{amsart}
\usepackage[T1]{fontenc}
\usepackage[latin9]{inputenc}
\usepackage{verbatim}
\usepackage{float}
\usepackage{amstext}
\usepackage{amsthm}
\usepackage{graphicx}

\makeatletter
\numberwithin{equation}{section}
\numberwithin{figure}{section}
\theoremstyle{plain}
\newtheorem{thm}{\protect\theoremname}
\theoremstyle{plain}
\newtheorem{prop}[thm]{\protect\propositionname}
\theoremstyle{remark}
\newtheorem{rem}[thm]{\protect\remarkname}

\usepackage{amsfonts}
\usepackage[font=small,labelfont=bf]{caption}
\usepackage[unicode=true,
 bookmarks=true,bookmarksnumbered=false,bookmarksopen=false,
 breaklinks=false,pdfborder={0 0 1},backref=false,colorlinks=false]
 {hyperref}

\makeatother

\usepackage{babel}
\providecommand{\propositionname}{Proposition}
\providecommand{\remarkname}{Remark}
\providecommand{\theoremname}{Theorem}

\begin{document}
\title{The exact renormalization group and dimensional regularization}
\author{Roberto Trinchero}
\date{05/11/2021}
\begin{abstract}
The exact renormalization group(ERG) is formulated implementing the
decimation of degrees of freedom by means of a particular momentum
integration measure. The definition of this measure involves a distribution
that links this decimation process with the dimensional regularization
technique employed in field theory calculations. Taking the dimension
$d=4-\epsilon$, the one loop solutions to the ERG equations for the
scalar field theory in this scheme are shown to coincide with the
dimensionally regularized perturbative field theory calculation in
the limit $\epsilon\to0^{-}$, if a particular relation between the
scale parameter $\mu$ and $\epsilon$ is employed. In general, it
is shown that in this scheme the solutions to the ERG equations for
the proper functions coincide when $\epsilon\to0^{-}$ with the complete
diagrammatic contributions appearing in field theory for these functions
and this theory, provided that exact relations between $\mu$ and
$\epsilon$ hold. In addition a non-perturbative approximation is
considered. This approximation consists in a truncation of the ERG
equations, which by means of a low momentum expansion leads to reasonable
results. 
\end{abstract}

\thanks{Supported by CONICET.}
\address{Instituto Balseiro and Centro Atómico Bariloche}
\maketitle

\section{Introduction}

The Wilson renormalization group(WRG)\cite{Wilson:1973jj} consists
in the study of the evolution of systems under scale transformations.
These scale transformations are implemented by means of the process
of decimation. Decimation consists in integrating over certain degrees
of freedom. A long distance effective theory is obtained by means
of integrating out short distance degrees of freedom. The separation
in short and long distance degrees of freedom involves the introduction
of a length scale. This separation can be done in many different ways.
In the original approach this was done by introducing a hard cut-off(HCO).
The HCO method consists in including a step function in the integration
measure for the modulus of the momenta. This procedure introduced
non-local terms in coordinate space\cite{Wegner-PhysRevA.8.401} and
therefore a soft cut-off was proposed\cite{Wilson:1973jj}. In the
soft cut-off (SCO)version the step function employed in the HCO is
replaced by a continuous function depending on a parameter which in
a certain limit for this parameter tends to the step function. The
implementation of this SCO into a effective field theory leads to
the ERG equations\footnote{Knowledge on this theme has been growing thanks to the contributions
of many authors. All of them will not be quoted here. Many references
on the subject can be found in the review \cite{bagnuls:hal-00012738}. }. These equations are exact, they can be written for the proper functions
of the theory, and take the form of one loop equations\cite{Weinberg1978}.
For the scalar $\phi$ field theory with symmetry under $\phi\to-\phi$,
these equations give the derivative respect to the scale parameter
of $n$-point proper functions in terms of other $j$-point functions,
with $j=2,\cdots,n+2$. 

On the other hand the field theoretic renormalization group(FTRG)\cite{stuckelberg-petermann1953normalisation}\cite{Gellmann-low-PhysRev.95.1300}
is related to making sense of the divergent contributions of Feynman
diagrams in perturbation theory. In particular from the requirement
that physical results should not depend on the subtraction procedure.
This requirement gives information about the behavior of $n$-point
functions under scaling of distances, which therefore provides a relation
with the WRG. In both approaches particular techniques have been developed
in order to study the renormalization group. In the FTRG dimensional
regularization\cite{bollini1972dimensional,tHooft:1972fi}(DR) and
minimal subtraction play a prominent role. They provide a very practical
calculation method and furthermore they are extremely useful when
dealing with gauge theories, since they preserve gauge symmetry. DR
is particular in the sense that the regularization is achieved by
means of a analytic continuation in the number of dimensions. This
is in contrast with other regularizations which directly modify the
high momentum behavior of the propagator. Another difference between
DR and other regularizations is that in DR it is necessary to introduce
two parameters, $\epsilon$ the dimensionless deviation from the integer
dimension $n$ considered and the energy scale $\mu$, which is used
to give couplings the correct dimension in $d=n-\epsilon$ dimensions.
In this work it will be shown that these two parameters should be
related in order to re-obtain the field theory results as the solution
to the ERG equations.

The purpose of this work is to give a version of the WRG which employs
dimensional regularization. In order to achieve this, the decimation
of degrees of freedom is linked with the dimensional regularization
technique. The features and results of this work are summarized as
follows, 
\begin{itemize}
\item A integration measure over high momenta degrees of freedom is given,
which links the HCO with dimensional regularization.
\item The ERG equations for the proper functions of the scalar field theory
with symmetry $\phi\to-\phi$ in four dimensions are considered in
the above mentioned scheme.
\item The $1$-loop perturbative solutions for the $2$-point and $4$-point
proper functions to the ERG equations are obtained. They are shown
to coincide with the dimensionally regularized field theoretic expressions,
when $\epsilon\to0^{-}$ and provided that a relation between the
scale parameter $\mu$ and $\epsilon$ holds\footnote{This relation is very similar to the one required to relate a perturbative
calculation done with dimensional regularization and the same calculation
done with a cut-off. }. Also the field renormalization function $\gamma$ is computed using
the corresponding ERG equation at two loops.
\item In general it is shown that the solutions to the ERG equations in
this scheme for the proper functions, coincide with the complete diagrammatic
contributions appearing in field theory for these functions when $\epsilon\to0^{-}$.
This is so, provided that for each $k$-point proper function a particular
relation between the scale parameter $\mu$ and $\epsilon$ holds.
This relation is exact and different for different $k$-point functions,
but not universal.
\item A non-perturbative approximation which consists in truncating the
ERG equations is considered. Supplementing this truncation with a
low momenta expansion gives to lowest order in the momenta reasonable
results for the running of the $2$-point and 4-point couplings. In
addition it is shown that higher orders in the momentum expansion
can be calculated using the usual results for the integrals appearing
in dimensionally regularized perturbative field theory. 
\end{itemize}
The paper is organized as follows. Section 2 presents the integration
measures and propagators of high and low momenta degrees of freedom.
Section 3 gives the ERG equations for proper functions as they appear
in ref. \cite{Weinberg1978}. Section 4 deals with a loop expansion
for the proper functions and computes the one loop expression for
the $2$-point, the $4$-point functions and the field renormalization
function $\gamma$ at two loops. Section 5 gives a diagrammatic non-perturbative
proof that the solutions of the ERG equations coincides with the expression
for these functions given in field theory. Section 6 presents a non-perturbative
approximation for the $2$ and $4$ point proper functions, which
consists in a truncation and a low momentum expansion. Finally, section
7 presents some concluding remarks and additional research motivated
by this work.

\section{Decimation and dimensional regularization}

The theory to be considered is described by the following functional
integral,
\begin{equation}
Z[J]=\int\mathcal{D}\phi\,e^{\left(-\frac{1}{2}\int\phi\Delta^{-1}\phi-S_{I}[\phi]+\int J\phi\right)}\label{eq:z}
\end{equation}
where the interaction term $S_{I}[\phi]$ is assumed to be given in
terms of polynomials in the fields and its first derivatives, and
such that,
\[
S[\phi]=S[-\phi]
\]
The propagator $\Delta$ appearing in (\ref{eq:z}) is taken to be\footnote{The dimension of space is taken to be $n$, however only the case
$n=4$ will be considered in this work.},
\begin{equation}
\Delta=\int\frac{d^{n}p}{(2\pi)^{n}}\Delta(p)\,e^{-ip\cdot(x-y)}\;\;,\Delta(p)=\frac{1}{p^{2}+m^{2}}\label{eq:delta-p}
\end{equation}

\subsection{Integration measures over high and low momenta}

The integration over momenta in $d$ dimensions is given by,
\begin{equation}
d^{d}p=d\Omega_{d-1}dp\,p^{d-1}\;\;\;,d=n-\epsilon\;\;,\label{eq:ddp}
\end{equation}
where $d\Omega_{d-1}$ is the integration measure over the $d-1$
dimensional unit sphere, for further use it is noted that,
\[
S_{d-1}=\int d\Omega_{d-1}=(2\pi)^{d/2}\Gamma(d/2)
\]
The integration over degrees of freedom is divided in two regions,
low momenta and high momenta. In the hard cut-off version, this is
done by using the following expression for the high momenta integration
measure $d_{H}^{n}p$,
\[
d_{H}^{n}p=d^{n}p\,\theta(\frac{p^{2}}{\mu^{2}}-1)
\]
where $\mu$ is the mass scale mentioned above, and gives the separation
between low momenta and high momenta. In this work this HCO is replaced
by a soft cut-off version implemented by means of,
\begin{align*}
(\theta(x-x_{0}),\phi(x)) & =\int_{x_{0}}^{\infty}dx\,\phi(x)\\
 & =\lim_{\epsilon\to0}(\theta_{\epsilon}(x-x_{0}),\phi(x))=\lim_{\epsilon\to0}\int_{x_{0}}^{\infty}dx\,x^{-\frac{\epsilon}{2}}\phi(x)
\end{align*}
in this respect the limit\footnote{It is emphasized that, following the same procedure as in dimensional
regularization, the limit $\epsilon\to0$ will be taken for the whole
expressions of the proper functions and not for each individual momentum
integration appearing in these expressions. This procedure is justified
by the results it leads to.} $\epsilon\to0$ is analogous to the hard cut-off limit mentioned
in \cite{Morris-94}. Decimation is implemented by integrating out
in (\ref{eq:z}) the high momenta degrees of freedom.

\subsection{The choice of the low $\Delta_{1}(p)$ and high $\Delta_{2}(p)$
momentum propagators}

The momentum space propagator $\Delta(p)$ is divided into two parts,
the low momenta propagator $\Delta_{1}$ and the high momenta propagator
$\Delta_{2}$, i.e.,
\begin{align*}
\Delta(p)= & \Delta_{1}(p,\mu)+\Delta_{2}(p,\mu)
\end{align*}
where,
\begin{align*}
\Delta_{2}(p,\mu) & =\frac{\theta_{\epsilon}(p^{2}/\mu^{2}-1)}{p^{2}+m^{2}}
\end{align*}
 The propagator will appear acting on functions of $p$, in expressions
of the form,
\[
I(q,\mu)=(\Delta_{2}(p,\mu),f(p,q))=\int d^{n}p\,\Delta_{2}(p,\mu)\,f(p,q)
\]
where $F(p,q)$ is in general given by a product of momentum space
complete propagators and $q$ generically denotes external momenta.
In appendix A the r.h.s. of the previous equation is written in terms
of a integration over $d=n-\epsilon$ dimensions and with no restriction
on the modulus of the momenta. Using equation (\ref{eq:thetaf}) in
appendix A for,
\[
f(p,q)=\frac{F(p.q)}{p^{2}+m^{2}}
\]
 leads to,
\begin{align}
I(q,\mu) & =\lim_{\epsilon\to0}\frac{S_{n-1}}{S_{d-1}}\int d^{d}p\mu^{\epsilon}\left\{ \frac{F(p.q)}{p^{2}+m^{2}}\right.\nonumber \\
 & \left.+\frac{\mu^{2}}{2p^{2}}\left[p\left(\frac{F^{(1,0)}(p,q)}{m^{2}+p^{2}}-\frac{2pF(p,q)}{\left(m^{2}+p^{2}\right)^{2}}\right)-2(\epsilon-1)\frac{F(p.q)}{p^{2}+m^{2}}\right]+\mathcal{O}(\frac{\mu^{4}}{p^{4}})\right\} \label{eq:delt2}
\end{align}
The action of $\mu\frac{\partial}{\partial\mu}\Delta_{2}(p)$ on functions
will also be employed in what follows, using equation (\ref{eq:dthetaf})
in appendix A, leads to,
\begin{align}
D(q,\mu) & =(\mu\frac{\partial}{\partial\mu}\Delta_{2}(p,\mu),F(p,q))\nonumber \\
 & =\epsilon I(q,\mu)+\lim_{\epsilon\to0}\frac{S_{n-1}}{S_{d-1}}\int d^{d}p\mu^{\epsilon}\times\nonumber \\
 & \times\left\{ \frac{\mu^{2}}{p^{2}}\left[p\left(\frac{F^{(1,0)}(p,q)}{m^{2}+p^{2}}-\frac{2pF(p,q)}{\left(m^{2}+p^{2}\right)^{2}}\right)-2(\epsilon-1)\frac{F(p.q)}{p^{2}+m^{2}}\right]+\mathcal{O}(\frac{\mu^{4}}{p^{4}})\right\} \label{eq:mudmuF}
\end{align}

\section{The ERG equations for proper functions }

The dimension in mass units of the momentum space dependent proper
function $\Gamma_{k}$ after factoring a momentum-conservation delta
function is\footnote{This means that,
\[
\bar{\Gamma}_{k}(p_{1},\cdots,p_{k})=\delta(p_{1}+\cdots+p_{k})\Gamma(p_{1},\cdots,p_{k})
\]
thus the arguments in $\Gamma$ should sum up to zero. The above relation
implies that,
\[
[\bar{\Gamma}_{k}]=[\Gamma_{k}]-d
\]
},
\begin{equation}
[\Gamma_{k}]=d-k\frac{(d-2)}{2}\overset{d=4-\epsilon}{=}4-\epsilon-k\left(1-\frac{\epsilon}{2}\right)=\left\{ \begin{array}{cc}
2 & for\,k=2\\
\epsilon & for\,k=4\\
-2+2\epsilon & for\,k=6\\
-4+3\epsilon & for\,k=8\\
\vdots & \vdots
\end{array}\right.\label{eq:dimgamma}
\end{equation}
The ERG equations arise as a consequence of the requirement that the
$k$ point correlators of the theory do not depend on the scale parameter\footnote{See \cite{Weinberg1978} p. 14.}
$\mu$. As shown in \cite{Weinberg1978}, this requirement leads to
the ERG equations satisfied by the proper functions\footnote{It is important to stress that the derivation of the ERG equations
appearing in \cite{Weinberg1978} , although diagrammatic, is non-perturbative
and exact. Indeed only a separation of the two point function into
two parts, which can always be done without any perturbative assumption,
and the fact that any correlator can be written as a tree graph with
vertices given by proper functions, are employed in this reference. } $\Gamma_{k}$, which are given by\footnote{In \cite{Weinberg1978} the ERG equations are written for the dimensionless
proper functions $\bar{g}_{k}=\mu^{-d+k\frac{(d-2)}{2}}\Gamma_{k}$
and in terms of the dimensionless momenta $l=p/\mu$. In this work
they are written in terms of dimension-full proper functions and momenta,
this is better suited for the aims of this work. },
\begin{align}
\left[\mu\frac{\partial}{\partial\mu}-p\cdot\frac{\partial}{\partial p}-k\frac{\gamma}{2}\right]\Gamma_{k}(p_{1},\cdots,p_{k}) & =L_{k}(\{p\};\mu)\label{eq:ergp}
\end{align}
where,
\begin{align}
L_{k}(p_{1},\cdots,p_{k};\mu) & =\frac{\mu^{-\epsilon}}{2}\int\frac{d^{n}p}{(2\pi)^{n}}\left(\mu\frac{\partial}{\partial\mu}\Delta_{2}(p;\mu)\right)\int\frac{d^{n}q_{1}}{(2\pi)^{n}}\cdots\frac{d^{n}q_{l-1}}{(2\pi)^{n}}\times\nonumber \\
 & \left[\sum_{p_{1}\cdots p_{l}}\bar{\Gamma}_{m_{1}+2}(p,-q_{1},p_{1}^{(1)},\cdots,p_{m_{1}}^{(1)})\mu^{-\epsilon}\Delta_{2}(q_{1})\right.\nonumber \\
 & \times\bar{\Gamma}_{m_{2}+2}(q_{1},-q_{2},p_{1}^{(2)},\cdots,p_{m_{2}}^{(2)})\mu^{-\epsilon}\Delta_{2}(q_{2})\cdots\nonumber \\
 & \left.\bar{\Gamma}_{m_{l}+2}(q_{l-1},-p,p_{1}^{(l)},\cdots,p_{m_{l}}^{(l)})\right]\label{eq:ln}
\end{align}
and,
\[
m_{1}+m_{2}+\cdots+m_{l}=k
\]
and the summation is over all possible ways of separating the $k$
momenta into $l$ sets, the set $j$ consisting of the $m_{j}$ momenta
$p_{1}^{(j)},\cdots,p_{m_{j}}^{(j)}$. 

\begin{figure}
\begin{centering}
\includegraphics[scale=0.3]{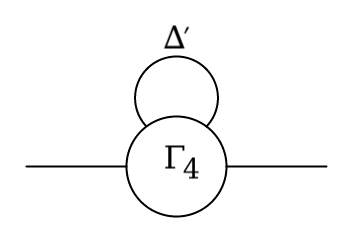}
\par\end{centering}
\caption{Graphical representation of $L_{2}$, in the figure $\Delta'=\mu\frac{\partial}{\partial\mu}\Delta_{2}$.\label{fig:The-one-loop-contribution l2}}

\end{figure}

\begin{figure}
\begin{centering}
\includegraphics[scale=0.3]{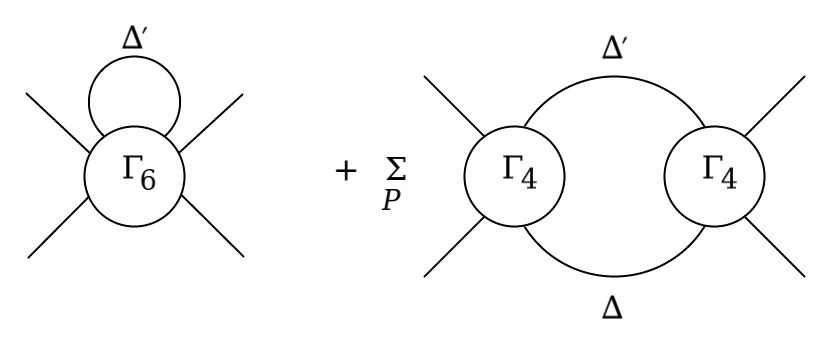}
\par\end{centering}
\caption{Graphical representation of $L_{4}$. In this figure $\sum_{P}$ indicates
summation over non-equivalent permutations of the external legs.\label{fig:The-one-loop-contribution-l4}}
\end{figure}

For $k=2$ the equation (\ref{eq:ergp}) gives,
\begin{align}
\mu\frac{\partial}{\partial\mu}\Gamma_{2}(p_{1};\mu) & =\left(\gamma+p_{1}\cdot\frac{\partial}{\partial p_{1}}\right)\Gamma_{2}(p_{1};\mu)+\nonumber \\
 & +\frac{\mu^{-\epsilon}}{2}\int\frac{d^{n}p}{(2\pi)^{n}}\left(\mu\frac{\partial}{\partial\mu}\Delta_{2}(p;\mu)\right)\Gamma_{4}(p,-p,p_{1},-p_{1};\mu)\label{eq:2p-full}
\end{align}
 For $k=4$, (\ref{eq:ergp}) leads to,
\begin{align}
\mu\frac{\partial}{\partial\mu}\Gamma_{4}(p_{1},p_{2},p_{3},p_{4};\mu) & =\left(2\gamma+\sum_{i=1}^{3}p_{i}\cdot\frac{\partial}{\partial p_{i}}\right)\Gamma_{4}(p_{1},p_{2},p_{3},p_{4};\mu)\nonumber \\
+\frac{\mu^{-\epsilon}}{2}\int\frac{d^{n}p}{(2\pi)^{n}} & \left(\mu\frac{\partial}{\partial\mu}\Delta_{2}(p;\mu)\right)\Gamma_{6}(p,-p,p_{1},p_{2},p_{3},p_{4};\mu)\nonumber \\
+\frac{\mu^{-\epsilon}}{2}\int\frac{d^{n}p}{(2\pi)^{n}}\frac{d^{n}p'}{(2\pi)^{n}} & \left(\mu\frac{\partial}{\partial\mu}\Delta_{2}(p;\mu)\Delta_{2}(p';\mu)\right)\nonumber \\
\times\left(\Gamma_{4}(p,-p',p_{1},p_{2};\mu)\right. & \,\Gamma_{4}(p',-p,p_{3},p_{4};\mu)\delta^{(n)}(p-p'+p_{1}+p_{2})\nonumber \\
+\Gamma_{4}(p,-p',p_{1},p_{3};\mu) & \Gamma_{4}(p',-p,p_{2},p_{4};\mu)\delta^{(n)}(p-p'+p_{1}+p_{3})\nonumber \\
+\Gamma_{4}(p,-p',p_{1},p_{4};\mu) & \left.\Gamma_{4}(p',-p,p_{2},p_{3};\mu)\delta^{(n)}(p-p'+p_{1}+p_{4})\right)\label{eq:4p-full}
\end{align}
the contributions involving momentum integrals to the right hand side
of the previous equations can be associated to the graphs in figs.
\ref{fig:The-one-loop-contribution l2} and \ref{fig:The-one-loop-contribution-l4}.

The initial conditions for a reference value $\mu_{0}$ of the mass
scale, are taken to be,
\begin{align}
\Gamma_{2}(p_{1};\mu_{0}) & =-\left(p_{1}^{2}+m_{I}^{2}\right)\;\;\;,\nonumber \\
\Gamma_{4}(p_{1},p_{2},p_{3};\mu_{0}) & =-\lambda_{I}\;\;\;\lambda_{I}\ll1,\nonumber \\
\Gamma_{2n}(\mu_{0}) & =0\;\;\forall\,n>2\label{eq:ic-1}
\end{align}
That is, the renormalization group flow is studied in the vicinity
of a point which lies in the $(2-4)$ plane in the space of couplings,
i.e. only $\Gamma_{2}$ and $\Gamma_{4}$ are assumed to be non-vanishing
at the scale $\mu_{0}$. The solutions to (\ref{eq:ergp}) depend
on the initial conditions (\ref{eq:ic-1}). It is noted that $\lambda_{I}$
is dimension-full and $g_{I}=\lambda_{I}\mu^{-\epsilon}$ dimensionless.

It is important to state that fields will be rescaled so that the
coupling $K$ corresponding to the kinetic term $Q$, 
\begin{equation}
Q=K\partial\phi\cdot\partial\phi\;\;,K=1\label{eq:k=00003D1}
\end{equation}
is equal to $1$. This entails a redefinition of the proper functions
so that the effective action,
\[
\Gamma[\phi]=\sum_{k=1}^{\infty}\frac{1}{k!}\int d^{n}x_{1}\cdots d^{n}x_{k}\,\Gamma_{k}(x_{1},\cdots,x_{k})\phi(x_{1})\cdots\phi(x_{k})
\]
remains the same. 

Condition (\ref{eq:k=00003D1}) in terms of the two point function
$G_{2}(p_{1},p_{2};\mu)$ is,
\begin{equation}
\left.\frac{\partial^{2}}{\partial p_{1}^{2}}\Gamma_{2}(p_{1},p_{2};\mu)\right|_{p_{1}=0}=-1\label{eq:kinetic-cond}
\end{equation}
The mass $m$ appearing in (\ref{eq:delta2}) and $m_{I}$ appearing
in this initial condition differ by the factor $\sqrt{Z(\mu)}$ that
accounts for the redefinition of the fields mentioned above, and is
given by,
\[
m^{2}=Z(\mu)m_{I}^{2}
\]
Taking the second derivative of equation (\ref{eq:2p-full}) respect
to the external momenta $p_{1}$, evaluating at $p_{1}=0$ and using
(\ref{eq:kinetic-cond})\footnote{It is worth noting that if other normalization of the kinetic term
is employed, such as $K'=\kappa$, then $\gamma$ is changed to $\gamma'=\frac{\gamma}{\kappa}$
.} , leads to a equation for $\gamma$,
\begin{equation}
\gamma=-\frac{\mu^{-\epsilon}}{2}\left.\frac{\partial^{2}}{\partial p_{1}^{2}}\int\frac{d^{n}p}{(2\pi)^{n}}\left(\mu\frac{\partial}{\partial\mu}\Delta_{2}(p;\mu)\right)\Gamma_{4}(p,-p,p_{1},-p_{1};\mu)\right|_{p_{1}=0}\label{eq:gamma}
\end{equation}

\section{Loop expansion\label{sec:Loop-expansion}}

A loop expansion for the proper functions consists in an expansion
of the form,
\begin{equation}
\Gamma_{2k}=\sum_{j}\gamma_{2k}^{(j)}\label{eq:pert}
\end{equation}
where $\gamma_{2k}^{(j)}$ is the contribution to $j$ loops. It is
also convenient to define the partial sums,
\[
\Gamma_{2k}^{(J)}=\sum_{j=1}^{J}\gamma_{2k}^{(j)}
\]
which give the $2k$-point function up to $J$ loops. 

It is important to bear in mind that in the r.h.s. of equations (\ref{eq:2p-full})
and (\ref{eq:4p-full}), the terms containing $\mu\frac{\partial}{\partial\mu}\Delta_{2}(p)$
are of one order higher in a loop expansion than the others. Thus
the equations for the loop expansion coefficients are,
\begin{align}
\mu\frac{\partial}{\partial\mu}\Gamma_{2}^{(j)}(p_{1};\mu) & =\left(\gamma+p_{1}\cdot\frac{\partial}{\partial p_{1}}\right)\Gamma_{2}^{(j)}(p_{1};\mu)+\nonumber \\
 & \frac{1}{2}\int\frac{d^{n}p}{(2\pi)^{n}}\left(\mu\frac{\partial}{\partial\mu}\Delta_{2}(p;\mu)\right)\Gamma_{4}^{(j-1)}(p,-p,p_{1},-p_{1};\mu)\label{eq:2p-loopj}
\end{align}
and,
\begin{align}
 & \mu\frac{\partial}{\partial\mu}\Gamma_{4}^{(j)}(p_{1},p_{2},p_{3},p_{4};\mu)=\left(2\gamma+\sum_{i=1}^{3}p_{i}\cdot\frac{\partial}{\partial p_{i}}\right)\Gamma_{4}^{(j)}(p_{1},p_{2},p_{3},p_{4};\mu)\nonumber \\
 & +\frac{1}{2}\int\frac{d^{n}p}{(2\pi)^{n}}\left(\mu\frac{\partial}{\partial\mu}\Delta_{2}(p;\mu)\right)\Gamma_{6}^{(j-1)}(p,-p,p_{1},p_{2},p_{3},p_{4};\mu)\nonumber \\
 & +\frac{1}{2}\int\frac{d^{n}p}{(2\pi)^{d}}d^{n}p'\left(\mu\frac{\partial}{\partial\mu}\Delta_{2}(p;\mu)\Delta_{2}(p';\mu)\right)\nonumber \\
 & \times\sum_{k,n/k+n=j-\text{1}}\left(\Gamma_{4}^{(k)}(p,-p',p_{1},p_{2};\mu)\,\Gamma_{4}^{(n)}(p',-p,p_{3},p_{4};\mu)\delta^{(d)}(p-p'+p_{1}+p_{2})\right.\nonumber \\
 & +\Gamma_{4}^{(k)}(p,-p',p_{1},p_{3};\mu)\Gamma_{4}^{(n)}(p',-p,p_{2},p_{4};\mu)\delta^{(d)}(p-p'+p_{1}+p_{3})\nonumber \\
 & \left.+\Gamma_{4}^{(k)}(p,-p',p_{1},p_{4};\mu)\Gamma_{4}^{(n)}(p',-p,p_{2},p_{3};\mu)\delta^{(d)}(p-p'+p_{1}+p_{4})\right)\label{eq:4p-loopj}
\end{align}
Also,
\[
\gamma^{(j)}=-\frac{\mu^{-\epsilon}}{2}\left.\frac{\partial^{2}}{\partial l_{1}^{2}}\int\frac{d^{n}l}{(2\pi)^{n}}\left(\mu\frac{\partial}{\partial\mu}\Delta_{2}(l;\mu)\right)\bar{g_{4}}^{(j-1)}(l,-l,l_{1},-l_{1};\mu)\right|_{l_{1}=0}
\]
It is worth remarking that the first contributions to $\gamma$ start
at two loops, because $\Gamma_{4}$ starts to depend on the external
momenta at one loop.

\subsection{k-point function at tree level }

The equation for $\Gamma_{k}^{(0)},\;k\,$ even is,
\[
\mu\frac{\partial}{\partial\mu}\Gamma_{k}^{(0)}(p_{1},\cdots,p_{k};\mu)=\left(\sum_{i=1}^{k}p_{i}\cdot\frac{\partial}{\partial p_{i}}\right)\Gamma_{k}^{(0)}(p_{1},\cdots,p_{k};\mu)
\]
whose general solution is,
\[
\Gamma_{k}^{(0)}(p_{1},\cdots,p_{k};\mu)=\Phi_{k}(\mu p_{1},\cdots,\mu p_{k})
\]
with the initial condition,
\[
\Gamma_{k}^{(0)}(p_{\text{1}},\cdots,p_{k};\mu_{0})=\begin{cases}
\begin{array}{c}
\Gamma_{2}^{(0)}(p_{1},p_{2};\mu_{0})=-\left(p_{1}^{2}+m_{I}^{2}\right)\;\;\;,\\
\Gamma_{4}^{(0)}(p_{1},p_{2},p_{3},p_{4};\mu_{0})=-\lambda_{I}\;\;\;\lambda_{I}\ll1,\\
0
\end{array} & \begin{array}{c}
k=2\\
k=4\\
k\geq6
\end{array}\end{cases}
\]
leads to the particular solution,
\[
\begin{array}{cc}
\Gamma_{2}^{(0)}(p_{1},p_{2};\mu)=-\left(p_{1}^{2}\left(\frac{\mu}{\mu_{0}}\right)^{2}+m_{I}^{2}\right) & k=2\\
\Gamma_{4}^{(0)}(p_{1},p_{2},p_{3},p_{4};\mu)=-\lambda_{I} & k=4\\
\Gamma_{k}^{(0)}(p_{1},\cdots,p_{k};\mu)=0 & k\geq6
\end{array}
\]
where $m_{I}$ and $\lambda_{I}$ are constants that do not depend
on $\mu$. Thus redefining the field by,
\[
\tilde{\phi}=Z(\mu)^{-\frac{1}{2}}\phi\;\;,\;Z(\mu)=\left(\frac{\mu_{0}}{\mu}\right)^{2}\Rightarrow m_{I}^{2}=\left(\frac{\mu}{\mu_{0}}\right)^{2}m^{2}
\]
which leads to a new k-point functions $\tilde{\Gamma}_{k}$, given
by,
\[
\tilde{\Gamma}_{k}=Z^{\frac{k}{2}}\Gamma_{k}
\]
in particular for $n=2$, this gives, 
\[
\tilde{\Gamma}_{2}(p_{1})=Z\Gamma_{2}(p_{1})=-\left(p_{1}^{2}+m^{2}\right)
\]

\subsection{Two point function at one loop}

\noindent To one loop order the above equation reduces to,
\begin{align*}
\left(\mu\frac{\partial}{\partial\mu}-p_{1}\cdot\frac{\partial}{\partial p_{1}}\right)\Gamma_{2}^{(1)}(p_{1};\mu)=\frac{1}{2}\int\frac{d^{n}p}{(2\pi)^{d}}\left(\mu\frac{\partial}{\partial\mu}\Delta_{2}(p)\right)\Gamma_{4}^{(0)}(p,-p,p_{1},-p_{1};\mu)
\end{align*}
this is so because $\gamma=0$ at this loop order. Also it is noted
that $\Gamma_{4}^{(0)}(p,-p,p_{1},-p_{1};\mu)=-\lambda_{I}$ independent
of momenta, thus,
\begin{equation}
\left(\mu\frac{\partial}{\partial\mu}-p_{1}\cdot\frac{\partial}{\partial p_{1}}\right)\Gamma_{2}^{(1)}(p_{1};\mu)=\lambda_{I}\mu^{-\epsilon}C(\mu)\label{eq:g21}
\end{equation}
where,
\begin{align*}
C(\mu) & =-\frac{1}{2}\int\frac{d^{n}p}{(2\pi)^{d}}\mu\frac{\partial}{\partial\mu}\Delta_{2}(p)
\end{align*}
employing equation (\ref{eq:mudmuF}) with $F(p,q)=1$, gives,
\begin{align}
C(\mu) & =-\frac{1}{2}\frac{S_{n-1}}{S_{d-1}}\int\frac{d^{d}p}{(2\pi)^{d}}\mu^{\epsilon}\left\{ \epsilon\frac{1}{p^{2}+m^{2}}+\frac{\mu^{2}}{p^{2}}.\left(\frac{\epsilon}{2}+1\right)\right.\label{eq:cmu}\\
 & \times\left.\left(-2\frac{p^{2}}{(p^{2}+m^{2})^{2}}+2\frac{(1-\epsilon)}{p^{2}+m^{2}}\right)+\mathcal{O}(\epsilon)+\mathcal{O}\left(\frac{\mu^{4}}{p^{4}}\right)\right\} 
\end{align}

\noindent it is noted that this expression is finite when $\epsilon\to0$.
In the first term inside the parenthesis this is so because this term
is multiplied by $\epsilon$. For the second term this is so because
the poles of the two terms inside the bracket cancel each other. The
above expression can be explicitly evaluated using the usual dimensional
regularization techniques leading to,
\[
C(\mu)=\frac{S_{n-1}}{S_{d-1}}\frac{m^{2}}{(4\pi)^{2}}\left(\frac{\mu}{m}\right)^{\epsilon}\left(1-\frac{\mu^{2}}{m^{2}}+\mathcal{O}(\epsilon)+\mathcal{O}\left(\frac{\mu^{4}}{p^{4}}\right)\right)
\]
 This last expression is proportional to the $\gamma_{m}$ function
defined by,
\begin{align}
\gamma_{m} & =\frac{\mu}{m^{2}}\frac{\partial}{\partial\mu}\Gamma_{2}^{(1)}(0;\mu)=\frac{S_{n-1}}{S_{d-1}}\frac{1}{m^{2}}\lambda_{I}\mu^{-\epsilon}\frac{m^{2}}{(4\pi)^{2}}\left(\frac{\mu}{m}\right)^{\epsilon}\left(1-\frac{\mu^{2}}{m^{2}}\right)\label{eq:gam}\\
 & =\frac{S_{n-1}}{S_{d-1}}\frac{g_{I}}{(4\pi)^{2}}\left(\frac{\mu}{m}\right)^{\epsilon}\left(1-\frac{\mu^{2}}{m^{2}}\right)\overset{\epsilon\to0}{=}\frac{g_{I}}{(4\pi)^{2}}\left(1-\frac{\mu^{2}}{m^{2}}\right)
\end{align}
it is important to note that the perturbative field theory beta function
is given by the first term in the parenthesis above. The second term
is a correction, which vanishes in the limit $\mu\to0$, i.e. when
the functional integration is done over all degrees of freedom.

\noindent Using the results in appendix B, the solution to (\ref{eq:g21})
which reduces to the zero loop result when neglecting one-loop corrections,
is\footnote{The $\sim$ over $\Gamma_{2}$ indicates that this is the rescaled
2-point proper function as defined in the previous subsection. },
\begin{align*}
\tilde{\Gamma}_{2}^{(1)}(p;\mu) & =-p^{2}-m^{2}-\lambda_{I}\left[\int_{r_{0}}^{r}\frac{d\tilde{r}}{\tilde{r}}\right]_{r=\mu^{-1}}C(\mu)\\
 & =-p^{2}-m^{2}-\lambda_{I}\left[\log(r/r_{0})\right]_{r=\mu^{-1}}C(\mu)\\
 & =-p^{2}-m^{2}+\lambda_{I}\log(\mu/\mu_{0})C(\mu)
\end{align*}

\noindent employing the expression in (\ref{eq:cmu}), the first term
inside the parenthesis, leads to, 
\begin{align}
\tilde{\Gamma}_{2}^{(1)}(p;\mu) & =-p^{2}-m^{2}-\frac{\lambda_{I}}{2}\frac{S_{n-1}}{S_{d-1}}\log\left(\frac{\mu}{\mu_{0}}\right)\epsilon\int\frac{d^{d}p}{(2\pi)^{d}}\frac{1}{p^{2}+m^{2}}+\mathcal{O}(\epsilon)+\mathcal{O}\left(\mu^{2}\right)\label{eq:gamma-2-1}
\end{align}
At this point, it is quite natural to ask the following relevant question,
how does the usual 1-loop perturbative expression for $\tilde{\Gamma}_{2}^{(1)}$
is obtained from this expression? The following relation between $\mu,\mu_{0}$
and $\epsilon$,
\begin{equation}
\frac{S_{n-1}}{S_{d-1}}\log\left(\frac{\mu}{\mu_{0}}\right)\epsilon=1\label{eq:mueps-1}
\end{equation}
gives an answer to the above question. Indeed, such a relation implies
that,
\begin{itemize}
\item The coefficient of the third term in (\ref{eq:gamma-2-1}) is exactly
the one appearing in the usual 1-loop perturbative expression for
$\tilde{\Gamma}_{2}^{(1)}$.
\item And,
\[
\frac{\mu}{\mu_{0}}=e^{\frac{1}{\epsilon}\frac{S_{n-1}}{S_{d-1}}}
\]
 thus, for $\epsilon\to0^{-}$ then $\mu\to0$, which therefore make
all the corrections $\mathcal{O}(\epsilon)$ and $\mathcal{O}\left(\mu^{2}\right)$
vanish in that limit.
\end{itemize}
\noindent It is worth noting that the limit $\mu\to0$ corresponds
to integrating over all degrees of freedom in the functional integral
in equation (\ref{eq:z}), which is just the procedure employed to
obtain the $1$-loop perturbative expression for $\Gamma_{2}^{(1)}$.
It should be stressed that the above relation (\ref{eq:mueps-1})
is only required if the intention is to re-obtain the field theoretic
result form the solution to ERG equation. 

\noindent This contribution corresponds to the tadpole diagram shown
below,
\begin{center}
\includegraphics[scale=0.3]{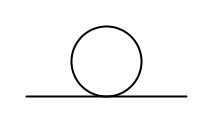}\captionof{figure}{Tadpole
diagram.}
\par\end{center}

\subsection{Four point function at one loop\label{subsec:Four-point-function}}

Equation (\ref{eq:4p-loopj}) for $j=0$ is,
\begin{align*}
\mu\frac{\partial}{\partial\mu}\Gamma_{4}^{(1)}(p_{1},\cdots,p_{4};\mu) & =\left(2\gamma+\sum_{i=1}^{3}p_{i}\cdot\frac{\partial}{\partial p_{i}}\right)\Gamma_{4}^{(1)}(p_{1},\cdots,p_{4};\mu)\\
+\frac{(-\lambda_{I})^{2}\mu^{-\epsilon}}{2}\int\frac{d^{n}p}{(2\pi)^{n}}\frac{d^{n}p'}{(2\pi)^{n}} & \left(\Delta_{2}(p)\mu\frac{\partial}{\partial\mu}\Delta_{2}(p')\right)\\
 & \sum_{P_{4}}\delta(p-p'+p_{i_{1}}+p_{i_{2}})
\end{align*}

\noindent where $\sum_{P_{4}}$ indicates a summation over all the
$\frac{4!}{2!2!}=6$ different permutations of the form,
\[
\left(\begin{array}{cccc}
1 & 2 & 3 & 4\\
i_{1} & i_{2} & i_{3} & i_{4}
\end{array}\right)
\]
to this loop order $\gamma=0$ and using that,
\[
\int d^{n}p\,\delta^{(n)}(p-p')F(p)=F(p')
\]
leads to,
\begin{align}
\left(\mu\frac{\partial}{\partial\mu}-\sum_{i=1}^{3}p_{i}\cdot\frac{\partial}{\partial p_{i}}\right)\Gamma_{4}^{(1)}(p_{1},\cdots,p_{4};\mu)= & \lambda_{I}^{2}\mu^{-\epsilon}A(p_{1},\cdots,p_{4})\label{eq:gral-4p-loop1}
\end{align}
where,
\begin{equation}
A(p_{1},\cdots,p_{4})=B(p_{1}+p_{2})+B(p_{1}+p_{3})+B(p_{1}+p_{4})\label{eq:a1234}
\end{equation}
and,
\begin{align*}
B(P) & =\frac{1}{2}\int\frac{d^{n}p}{(2\pi)^{n}}\left(\mu\frac{\partial}{\partial\mu}\Delta_{2}(p)\Delta_{2}(p+P)\right)\\
 & =\frac{1}{2}\int\frac{d^{d}p}{(2\pi)^{d}}\left(\frac{\mu^{\epsilon}}{p^{2}+m^{2}}+\mathcal{O}\left(\frac{\mu^{2}}{p^{4}}\right)\right)\left(\frac{\epsilon}{(p+P)^{2}+m^{2}}+\mathcal{O}\left(\frac{\mu^{2}}{(p+P)^{4}}\right)\right)\\
 & =\frac{1}{2}\int\frac{d^{d}p}{(2\pi)^{d}}\frac{\epsilon\mu^{\epsilon}}{(p^{2}+m^{2})((p+P)^{2}+m^{2})}+\epsilon\times K+\mathcal{O}(\mu^{2})
\end{align*}
where $K$ denotes a quantity which is finite when $\epsilon\to0$.
The solution to (\ref{eq:gral-4p-loop1}) which reduces to the zero
loop result when neglecting one-loop corrections, is,
\begin{align*}
\Gamma_{4}^{(1)}(p_{1},\cdots,p_{4};\mu) & =-\lambda_{I}-\lambda_{I}^{2}\mu^{-\epsilon}\left[\int_{r_{0}}^{r}\frac{d\tilde{r}}{\tilde{r}}A(\tilde{r}\mu\{p\})\right]_{r=\mu^{-1}}
\end{align*}

\noindent in order to proceed the quantity $A(\tilde{r}\mu\{p\})$
is expanded as a power series in its argument. Only the first term
in that expansion, i.e. $A(0)$ can contribute to $\Gamma_{4}^{(1)}$,
this is so because the integrals appearing in $B(P)$ are multiplied
by $\epsilon$, and only the first term in that power series can give
something proportional to $\epsilon^{-1}$, the higher terms converge.
Thus the relevant integral over $\tilde{r}$ is,
\[
\int_{r_{0}}^{r}\frac{d\tilde{r}}{\tilde{r}}=\log(r/r_{0})
\]
 Thus,
\begin{align*}
\Gamma_{4}^{(1)}(p_{1},\cdots,p_{4};\mu) & =-\lambda_{I}+\lambda_{I}^{2}\log(\mu/\mu_{0})\frac{\epsilon}{2}\frac{S_{n-1}}{S_{d-1}}\times\\
 & \int\frac{d^{d}p}{(2\pi)^{d}}\sum_{P}\frac{1}{(p^{2}+m^{2})((p+P)^{2}+m^{2})}+\epsilon\times K+\mathcal{O}(\mu^{2})\\
 & =-\lambda_{I}+\lambda_{I}^{2}\frac{1}{2}\sum_{P}\int\frac{d^{d}p}{(2\pi)^{d}}\frac{1}{(p^{2}+m^{2})((p+P)^{2}+m^{2})}+\epsilon\times K+\mathcal{O}(e^{-\frac{1}{2|\epsilon|}\frac{S_{n-1}}{S_{d-1}}})
\end{align*}
where $\sum_{P}$ indicates a summation over $P=\{p_{1}+p_{2},p_{1}+p_{3},p_{1}+p_{4}\}$.
As happens for the case of the two point function, relation (\ref{eq:mueps-1})
makes the term proportional to $\log(\mu/\mu_{0})\epsilon$ give the
well known dimensionally regularized 1-loop expression for this correction
and at the same time, makes the $\mathcal{O}(\mu^{2})$ corrections
be negligible in the hard cut-off limit $\epsilon\to0^{-}$. Thus
assuming relation (\ref{eq:mueps-1}), then the dimensionally regularized
1-loop expression is re-obtained as the solution of the 1-loop ERG
equation. This contribution corresponds to the fish diagram shown
below,
\begin{center}
\includegraphics[scale=0.2]{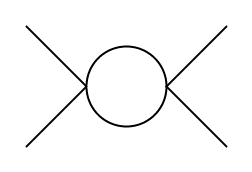}\captionof{figure}{Fish diagram.}
\par\end{center}

\noindent 

\subsection{$\gamma$ at two loops}

The equation to be considered is,
\begin{align*}
\gamma^{(2)} & =-\frac{\mu^{-\epsilon}}{2}\left.\frac{\partial^{2}}{\partial p_{1}^{2}}\int\frac{d^{d}p}{(2\pi)^{d}}\left(\mu\frac{\partial}{\partial\mu}\Delta_{2}(p;\mu)\right)\Gamma_{4}^{(1)}(p,-p,p_{1},-p_{1};\mu)\right|_{p_{1}=0}\\
 & =-\frac{\partial^{2}}{\partial p_{1}^{2}}\int\frac{d^{d}p}{(2\pi)^{d}}\left(\frac{\epsilon}{p^{2}+m^{2}}+\epsilon\times K'+\mathcal{O}(\mu^{2})\right)\times\\
 & \;\;\left.\times\left(\lambda_{I}^{2}\log(\mu/\mu_{0})\frac{\epsilon}{2}\frac{S_{n-1}}{S_{d-1}}\int\frac{d^{d}p'}{(2\pi)^{d}}\frac{1}{(p'^{2}+m^{2})((p+p'+p_{1})^{2}+m^{2})}+\epsilon\times K+\mathcal{O}(\mu^{2})\right)\right|_{p_{1}=0}\\
 & =-\frac{1}{2}\lambda_{I}^{2}\left.\frac{\partial^{2}}{\partial p_{1}^{2}}\int\frac{d^{d}p}{(2\pi)^{d}}\int\frac{d^{d}p'}{(2\pi)^{d}}\left(\epsilon\frac{1}{p^{2}+m^{2}}\frac{1}{(p'^{2}+m^{2})((p+p'+p_{1})^{2}+m^{2})}\right)\right|_{p_{1}=0}\\
 & +\epsilon\times K"+\mathcal{O}(e^{-\frac{1}{2|\epsilon|}\frac{S_{n-1}}{S_{d-1}}})
\end{align*}
where in the second equality it was used that the first term in (\ref{eq:a1234})
does not contribute to $\gamma$ because the corresponding integral
does not depend on $p_{1}$, so that upon taking the derivative respect
to this variable gives $0$. The other terms give the same contribution
by pairs. In the third equality, relation (\ref{eq:mueps-1}) was
employed. The first term in the r.h.s. of the last equality corresponds
to the sunrise  diagram, shown below,
\noindent \begin{center}
\includegraphics[scale=0.3]{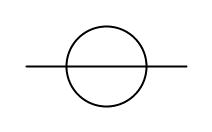}\captionof{figure}{Sunrise
diagram.} 
\par\end{center}

\noindent \begin{flushleft}
thus it leads to the usual field theoretic two loop contribution to
the $\gamma$ function associated with field renormalization. The
$\epsilon$ in front of this expression retains only the coefficient
of the simple pole when $\epsilon\to0$ of the second derivative of
this integral. 
\par\end{flushleft}

\section{Solutions of dimensionally regularized ERG equations coincide with
field theory calculation when $\epsilon\to0^{-}$\label{sec:Solutions-of-dimensionally}}

The general form of the dimensionally regularized ERG equation is,
\[
\left[\mu\frac{\partial}{\partial\mu}-p\cdot\frac{\partial}{\partial p}-\frac{n\gamma}{2}\right]\Gamma_{k}(\{p\};\mu)=L_{k}(\{p\};\mu)
\]
where $\{p\}$ denotes the whole set of the $k$ external momenta
$p_{1},p_{2},\cdots,p_{k}$ on which $\Gamma_{k}$ depends. The general
solution to this equation, as shown in Appendix B, is given by,
\[
\Gamma_{k}(\{p\},m,\mu)=\mu^{k\frac{\gamma}{2}}\left(\Phi(\mu\{p\},m)-\left[\int_{r_{0}}^{r}\frac{d\tilde{r}}{\tilde{r}^{1-k\frac{\gamma}{2}}}L_{k}(\tilde{r}\mu\{p\},m,\mu)\right]_{r=\mu^{-1}}\right)
\]
where $\Phi(p\mu)$ is an arbitrary function of $p\mu$ to be fixed
by boundary conditions. In this section it is shown that the above
solution to the ERG equations coincides with the expression of the
$n$-point proper function in terms of proper vertices linked by $\frac{1}{p^{2}+m^{2}}$
propagators when $\epsilon\to0^{-}$.

First the case of the two point function $\Gamma_{2}$ is considered.
The ERG equation in this case is, 
\begin{equation}
\left(\mu\frac{\partial}{\partial\mu}-\gamma-p_{1}\cdot\frac{\partial}{\partial p_{1}}\right)\Gamma_{2}(p_{1};\mu)=L_{2}(\,p_{1},\mu)\label{eq:gr2}
\end{equation}
where,
\[
L_{2}(\,p_{1},\mu)=\frac{\mu^{-\epsilon}}{2}\int\frac{d^{n}p}{(2\pi)^{n}}\left(\mu\frac{\partial}{\partial\mu}\Delta_{2}(p;\mu)\right)\Gamma_{4}(p,-p,p_{1},-p_{1};\mu)
\]
using (\ref{eq:mudmuF}) leads to the following expression for $L_{2}$,
\begin{equation}
L_{2}(\,p_{1},\mu)=\frac{1}{2}\frac{S_{n-1}}{S_{d-1}}\int\frac{d^{d}p}{(2\pi)^{.d}}\frac{\epsilon}{p^{2}+m^{2}}\Gamma_{4}(p,-p,p_{1},-p_{1};\mu)+\mathcal{O}\left(\frac{\mu^{2}}{p^{2}}\right)\label{eq:muddelta2}
\end{equation}
the first term on the r.h.s. of the last equation is denoted by $L_{2}^{RD}(\,p_{1},\mu)$.
In what follows the second terms of $\mathcal{O}\left(\mu^{2}/p^{2}\right)$
will be neglected, this procedure will be justified a posteriori.
As shown in appendix B, the solution to (\ref{eq:gr2}) is,
\begin{equation}
\Gamma_{2}(p_{1},m,\mu)=\mu^{\gamma}\left(\Phi(\mu p_{1},m)-\left[\int_{r_{0}}^{r}\frac{d\tilde{r}}{\tilde{r}^{1-\gamma}}L_{2}^{RD}(\tilde{r}\mu p_{1},m,\mu)\right]_{r=\mu^{-1}}\right)\label{eq:soll2}
\end{equation}
in order to perform the integration appearing in the second term of
the r.h.s. in the previous equation it is convenient to expand $L_{2}^{RD}(\tilde{r}\mu p_{1},m,\mu)$
in powers of the first argument, that is,
\[
L_{2}^{RD}(\tilde{r}\mu p_{1},m,\mu)=\frac{1}{2}\frac{S_{n-1}}{S_{d-1}}\int\frac{d^{d}p}{(2\pi)^{.d}}\frac{\epsilon}{p^{2}+m^{2}}\sum_{j}\frac{(\tilde{r}^{2}\mu^{2}p_{1}^{2})^{j}}{2j!}\left[\frac{\partial^{(j)}}{\partial^{j}p_{1}^{2}}\Gamma_{4}(p,-p,p_{1},-p_{1};\mu)\right]{}_{p_{1}=0}
\]
replacing this expression in (\ref{eq:soll2}), the integrals over
$\tilde{r}$ are of the general form,
\begin{align}
\mu^{2j+\gamma}\left[\int_{r_{0}}^{r}d\tilde{r}\tilde{r}^{\gamma-1+2j}\right]_{r=\mu^{-1}} & =\mu^{2j+\gamma}\left(\frac{r^{\gamma+2j}-r_{o}^{2\gamma+j}}{\gamma+2j}\right)_{r=\mu^{-1}}\nonumber \\
 & =\mu^{2j+\gamma}\left(\frac{\mu^{-\gamma-2j}-\mu_{o}^{-\gamma-2j}}{\gamma+2j}\right)\nonumber \\
 & =\left(\frac{1-(\mu/\mu_{0})^{\gamma+2j}}{\gamma+2j}\right)\label{eq:intrm}
\end{align}
At this point , it is important to note that, all the integrals over
the momentum $p$ appearing in $L_{2}^{RD}$ are convergent for the
terms in the summation with $j>1$. Thus, those terms do not contribute
to $L_{2}^{RD}$ in the limit $\epsilon\to0$. Noting the boundary
condition (\ref{eq:ic-1}), i.e. that for $\mu=\mu_{0}$, $\Gamma_{2}=-p_{1}^{2}-m_{I}^{2}$,
leads to,
\[
\Phi(\mu p_{1},m)=-\left(p_{1}\frac{\mu}{\mu_{0}}\right)^{2-\gamma}-\left(m\frac{\mu}{\mu_{0}}\right)^{2-\gamma}
\]
 Thus the term with $j=0$ gives the following contribution to $\Gamma_{2}$,
\[
\Gamma_{2}(p_{1},m,\mu)=-p_{1}^{2}-m^{2}-\frac{\epsilon}{\gamma}\frac{S_{n-1}}{S_{d-1}}\left(1-(\mu/\mu_{0})^{\gamma}\right)\frac{1}{2}\int\frac{d^{d}p}{(2\pi)^{d}}\frac{1}{p^{2}+m^{2}}\Gamma_{4}(p,-p,0,0)
\]
therefore if the following relation between $\epsilon$ and $\mu$
is imposed\footnote{Although in general $\gamma\neq0$, in the $1$-loop perturbative
approximation it vanishes and the integral over $r$ above for $\gamma=0$
gives the relation,
\[
\frac{1}{\epsilon}=\frac{S_{n-1}}{S_{d-1}}\log(\mu/\mu_{0})=\lim_{\gamma\to0}-\frac{1}{\gamma}\frac{S_{n-1}}{S_{d-1}}\left(1-(\mu/\mu_{0})^{\gamma}\right)
\]
which is the one that was employed in the $1$-loop computations in
section \ref{sec:Loop-expansion}.},
\begin{equation}
1=-\frac{\epsilon}{\gamma}\frac{S_{n-1}}{S_{d-1}}\left(1-(\mu/\mu_{0})^{\gamma}\right)\label{eq:mu-eps-gamma}
\end{equation}
then,
\begin{align}
\Gamma_{2}(p_{1}.m,\mu) & =-p_{1}^{2}-m^{2}+\frac{1}{2}\int\frac{d^{d}p}{(2\pi)^{d}}\frac{1}{p^{2}+m^{2}}\Gamma_{4}(p,-p,0,0)\label{eq:g21-1-1}
\end{align}
the r.h.s. of (\ref{eq:g21-1-1}) is the dimensionally regularized
expression for the proper two point function. Relation (\ref{eq:mu-eps-gamma})
is required in order to obtain the $2$-point function computed with
dimensional regularization as a solution of the ERG equations. This
relation links, the two parameters introduced by dimensional regularization,
the dimensionless $\epsilon$ and the energy scale $\mu$. According
to this relation, the limits $\mu\to0$ and $\epsilon\to0^{-}$ are
equivalent. Relation (\ref{eq:mu-eps-gamma}) implies that,
\[
\frac{\mu}{\mu_{0}}=\left(1+\frac{\gamma}{\epsilon}\frac{S_{d-1}}{S_{n-1}}\right)^{1/\gamma}\overset{\epsilon\to0^{-}}{=}e^{-\frac{1}{|\epsilon|}\frac{S_{d-1}}{S_{n-1}}}
\]
showing that the $\mathcal{O}(\mu^{2}/p^{2})$ appearing on the r.h.s.
of (\ref{eq:muddelta2}) can be safely neglected in the limit $\epsilon\to0^{-}$,
as was done above. Inserting the loop expansion (\ref{eq:pert}) for
$\Gamma_{4}(p,-p,p_{1},-p_{1})$ leads to a loop expansion for $\Gamma_{2}(l_{1};\mu).$ 

The term with $j=1$ is the coefficient of the kinetic term. As stated
in (\ref{eq:kinetic-cond}) this coefficient was fixed to $1$. Thus
the contribution of this term is already taking into account in $\gamma$
given by (\ref{eq:gamma}).

Next, the general case of $\Gamma_{k}$ is considered. Appendix B
gives the following general solution to the ERG equation for the $k$-point
proper function,
\begin{equation}
\Gamma_{k}(\{p\},m,\mu)=\mu^{k\frac{\gamma}{2}}\left(\Phi(\mu\{p\},m)-\left[\int_{r_{0}}^{r}\frac{d\tilde{r}}{\tilde{r}^{1-k\frac{\gamma}{2}}}L_{k}(\tilde{r}\mu\{p\},m,\mu)\right]_{r=\mu^{-1}}\right)\label{eq:gnmu}
\end{equation}
 with $L_{k}(\{p\},m,\mu)$ given by,
\begin{align}
L_{k}(p_{1},\cdots,p_{k},m,\mu) & =\frac{\mu^{-\epsilon}}{2}\int\frac{d^{d}p}{(2\pi)^{d}}\left(\frac{\epsilon}{p^{2}+m^{2}}+\mathcal{O}(\mu^{2}/p^{2})\right)\left[\sum_{l=1}^{\infty}d^{d}q_{1}\cdots d^{d}q_{l-1}\right.\nonumber \\
 & \times\sum_{m_{1}\cdots m_{l}}\bar{\Gamma}_{m_{1}+2}(p,-q_{1},p_{1}^{(1)},\cdots,p_{m_{1}}^{(1)})\mu^{-\epsilon}\left(\frac{\mu^{\epsilon}}{q_{1}^{2}+m^{2}}+\mathcal{O}(\mu^{2}/q_{1}^{2})\right)\nonumber \\
 & \times\bar{\Gamma}_{m_{2}+2}(q_{1},-q_{2},p_{1}^{(2)},\cdots,p_{2}^{(2)})\mu^{-\epsilon}\left(\frac{\mu^{\epsilon}}{q_{2}^{2}+m^{2}}+\mathcal{O}(\mu^{2}/q_{2}^{2})\right)\nonumber \\
 & \left.\times\cdots\bar{\Gamma}_{m_{l}+2}(q_{k-1},-p,p_{1}^{(l)},\cdots,p_{m_{l}}^{(l)})\right]\label{eq:ln-1}
\end{align}
where,
\[
m_{1}+m_{2}+\cdots+m_{l}=k
\]
and the summation is over all possible ways of separating the $k$
momenta into $l$ sets, the set $j$ consisting of the $m_{j}$ momenta
$p_{1}^{(j)},\cdots,p_{m_{j}}^{(j)}$. In (\ref{eq:ln-1}), equations
(\ref{eq:delt2}) and (\ref{eq:mudmuF}) were employed to write $\Delta_{2}$
and $\mu\frac{\partial}{\partial\mu}\Delta_{2}$ respectively. Denoting
by $L_{k}^{RD}(p_{1},\cdots,p_{k},m,\mu)$ the contribution to $L_{k}$
obtained by neglecting the $\mathcal{O}\left(\mu^{2}/p^{2}\right)$
terms , leads to,
\begin{align*}
\bar{\Gamma}_{k}(\{p\};\mu) & =\mu^{\frac{k}{2}\gamma}\left(\Phi(\{p\}\mu,m)-\int_{\mu_{0}}^{\mu}d\tilde{r}\,\tilde{r}^{-1+\frac{k}{2}\gamma}L_{k}^{RD}(\tilde{r}\mu\{p\},m,\mu)\right)
\end{align*}
proceeding as in the case of the $2$-point function, that is expanding
$L_{k}^{RD}(\tilde{r}\mu\{p\},m,\mu)$ as a power series in the first
argument, the integrals over $\tilde{r}$ are the same as in (\ref{eq:intrm}).
For $k>2$ only the first term of the power series mentioned above
contributes in the limit $\epsilon\to0$. Extracting the factor $\frac{S_{n-1}}{S_{d-1}}\epsilon$
in front of $L_{k}(p)$ and using the relation,
\begin{equation}
\frac{1}{\epsilon}=-\frac{2}{k\gamma}\frac{S_{n-1}}{S_{d-1}}\left(1-(\mu/\mu_{0})^{\frac{k}{2}\gamma}\right)\label{eq:mu-eps-gammak}
\end{equation}
 gives,
\begin{equation}
\Gamma_{k}(\{p\},m,\mu)=\mu^{\frac{k}{2}\gamma}\left(\Phi(\{p\}\mu,m\})-\tilde{\Gamma}_{k}^{RD}(\{p\},m,\mu)\right)\label{eq:gamma-n}
\end{equation}
where $\tilde{\Gamma}_{k}^{RD}(p)$ is given by,
\begin{align}
\tilde{\Gamma}_{k}^{RD}(\{p\},m,\mu)= & \frac{\mu^{-\epsilon}}{2}\int\frac{d^{d}p}{(2\pi)^{d}}\frac{\epsilon}{p^{2}+m^{2}}\left[\sum_{l=1}^{\infty}d^{d}p_{1}\cdots d^{d}p_{l}\right.\nonumber \\
 & \times\sum_{m_{1}\cdots m_{l}}\bar{\Gamma}_{m_{1}+2}(p,-p_{1},p_{1}^{(1)},\cdots,p_{m_{1}}^{(1)})\frac{1}{p_{1}^{2}+m^{2}}\nonumber \\
 & \times\bar{\Gamma}_{m_{2}+2}(p_{1},-p_{2},p_{1}^{(2)},\cdots,p_{2}^{(2)})\frac{1}{p_{2}^{2}+m^{2}}\nonumber \\
 & \left.\times\cdots\bar{\Gamma}_{m_{l}+2}(p_{k-1},-p,p_{1}^{(l)},\cdots,p_{m_{l}}^{(l)})\right]\label{eq:Lmo=0000F1o}
\end{align}
as for the case of $L_{2}$, the contribution of the additional $\mathcal{O}\left(\mu^{2}/p^{2}\right)$
appearing in (\ref{eq:ln-1}) can be safely neglected in the limit
$\epsilon\to0^{-},$ if relation (\ref{eq:mu-eps-gammak}) is imposed.
In appendix C it is shown that expression (\ref{eq:gamma-n}) coincides
with the one obtained in field theory. 

It is important to emphasize that the fact that the relation,
\begin{equation}
\frac{1}{\epsilon}=-\frac{S_{n-1}}{S_{d-1}}\frac{2}{k\gamma}\left(1-(\mu/\mu_{0})^{\frac{k}{2}\gamma}\right)\label{eq:k-mu-eps}
\end{equation}
is required to obtain coincidence of the solutions of the ERG equations
for the $k$-point proper function with the field theory contributions,
is a non-perturbative exact result. This is so since in this section
no perturbative approximation has been employed. On the other hand
this relation is not universal, indeed all critical exponents are
given in terms of beta functions, which do not involve the integrals
over $r$ which give rise to relation (\ref{eq:k-mu-eps}). In addition
if $\epsilon$ is changed by a factor, then this factor does not affect
the critical exponents, but it will change the relation (\ref{eq:k-mu-eps}),
also if the normalization of the kinetic term is changed then, as
shown above, $\gamma$ changes and this affects the form of relation
(\ref{eq:k-mu-eps}). 

\section{Non-perturbative approximation }

The approximation consists in first truncating the ERG to a given
order. The truncation where $\Gamma_{n}=0\;,\forall n>4$ is considered.
This truncation is justified since $\Gamma_{n}$ for $n>4$ correspond
to irrelevant operators whose couplings will vanish when going to
the infrared. Then the truncated equations for $\Gamma_{2}$ and $\Gamma_{4}$
are (\ref{eq:2p-full}) and (\ref{eq:4p-full}), where in the last
one, due to the truncation, the first term on the r.h.s. does not
appear, i.e. the truncated ERG equations are,
\begin{align*}
\mu\frac{\partial}{\partial\mu}\Gamma_{2}(p_{1};\mu) & =\left(\gamma+p_{1}\cdot\frac{\partial}{\partial p_{1}}\right)\Gamma_{2}(p_{1};\mu)\\
 & +\frac{\mu^{-\epsilon}}{2}\int\frac{d^{d}p}{(2\pi)^{d}}\left(\mu\frac{\partial}{\partial\mu}\Delta_{2}(p;\mu)\right)\Gamma_{4}(p,-p,p_{1},-p_{1};\mu)
\end{align*}

\begin{align}
\mu\frac{\partial}{\partial\mu}\Gamma_{4}(p_{1},p_{2},p_{3},p_{4};\mu) & =\left(2\gamma+\sum_{i=1}^{4}p_{i}\cdot\frac{\partial}{\partial p_{i}}\right)\Gamma_{4}(p_{1},p_{2},p_{3},p_{4};\mu)\nonumber \\
 & +\frac{\mu^{-2\epsilon}}{2}\int\frac{d^{d}p}{(2\pi)^{d}}d^{d}p'\left(\Delta_{2}(p;\mu)\mu\frac{\partial}{\partial\mu}\Delta_{2}(p';\mu)\right)\nonumber \\
 & \times\left(\Gamma_{4}(p,-p',p_{1},p_{2};\mu)\Gamma_{4}(p',-p,p_{3},p_{4};\mu)\delta^{(d)}(p-p'+p_{1}+p_{2})\right.\nonumber \\
 & +\Gamma_{4}(p,-p',p_{1},p_{3};\mu)\Gamma_{4}(p',-p,p_{2},p_{4};\mu)\delta^{(d)}(p-p'+p_{1}+p_{3})\nonumber \\
 & \left.+\Gamma_{4}(p,-p',p_{1},p_{4};\mu)\Gamma_{4}(p',-p,p_{2},p_{3};\mu)\delta^{(d)}(p-p'+p_{1}+p_{4})\right)\label{eq:4p-full-1}
\end{align}
as in section \ref{sec:Solutions-of-dimensionally} , the derivative
respect to the external momenta in these equations can be written
as,
\[
\sum_{i=1}^{k}p_{i}\cdot\frac{\partial}{\partial p_{i}}\Gamma_{k}(p_{1},\cdots,p_{k})=\left.\rho\frac{\partial}{\partial\rho}\Gamma_{k}(\rho p_{1},\cdots,\rho p_{k})\right|_{\rho=1}
\]
in addition the following low momenta expansion is proposed,
\[
\Gamma_{k}(\rho p_{1},\cdots,\rho p_{k})=\sum_{j=0}^{\infty}\rho^{j}\Gamma_{k}^{(j)}(p_{1},\cdots,p_{k})
\]
putting $\rho=0$ on both sides of this equation implies that $\Gamma_{k}^{(0)}$
does not depend on the momenta, in general the coefficient $\Gamma_{k}^{(j)}(p_{1},\cdots,p_{k})$,
will be a polynomial of degree $j$ in the momenta with all Lorentz
indices contracted. In addition the zeroth order $\gamma$ function
given by (\ref{eq:gamma}) vanishes, because $\Gamma_{4}^{(0)}$ does
not depend on the external momenta. Thus the truncated equation up
to order $\mathcal{O}(\rho^{0})$ is,
\begin{align*}
\mu\frac{\partial}{\partial\mu}\Gamma_{4}^{(0)}(\mu) & =\frac{3}{2}\left(\Gamma_{4}^{(0)2}(\mu)\mu^{-2\epsilon}\right)\frac{1}{2}\int\frac{d^{d}p}{(2\pi)^{d}}\left(\Delta_{2}(p;\mu)\mu\frac{\partial}{\partial\mu}\Delta_{2}(p;\mu)\right)\\
 & =\frac{3}{4}\left(\Gamma_{4}^{(0)2}(\mu)\mu^{-2\epsilon}\right)\epsilon\int\frac{d^{d}p}{(2\pi)^{d}}\left(\frac{\mu^{2\epsilon}}{(p^{2}+\Gamma_{2}^{(0)}(\mu)^{2})^{2}}\right)\\
 & =\frac{3}{4}\Gamma_{4}^{(0)2}(\mu)\frac{\Gamma_{2}^{(0)}(\mu)^{-\epsilon}}{(4\pi)^{2-\frac{\epsilon}{2}}}\epsilon\Gamma(\frac{\epsilon}{2})\overset{\epsilon\to0}{=}\frac{3}{2(4\pi)^{2}}\Gamma_{4}^{(0)2}(\mu)
\end{align*}
the previous equation gives the beta function for the coupling $\Gamma_{4}^{(0)}(\mu)$.
The solution to this equation is,
\[
\Gamma_{4}^{(0)}(\mu)=-\frac{\lambda_{I}}{1-\lambda_{I}\frac{3}{2(4\pi)^{2}}\log\left(\frac{\mu}{\mu_{0}}\right)}
\]
, which satisfies the boundary condition (\ref{eq:ic-1}). This shows
that the effective coupling $\lambda=-\Gamma_{4}^{(0)}(\mu)$ as a
function of $\mu$ for $\mu<\mu_{0}$, is smaller than its initial
value $\lambda_{I}$. Even if the initial coupling $\lambda_{I}$
is $\lambda_{I}\gg1$, the effective coupling saturates at a value
$\lambda_{s}=-\frac{(4\pi)^{2}}{3\log\left(\frac{\mu}{\mu_{0}}\right)}$
. Replacing this result in the equation for $\Gamma_{2}^{(0)}(\mu)$
leads to,
\begin{align*}
\mu\frac{\partial}{\partial\mu}\Gamma_{2}^{(0)}(\mu) & =-\frac{\mu^{-\epsilon}}{2}\frac{\lambda_{I}}{1-\lambda_{I}\frac{3}{2(4\pi)^{2}}\log\left(\frac{\mu}{\mu_{0}}\right)}\int\frac{d^{d}p}{(2\pi)^{d}}\left(\mu\frac{\partial}{\partial\mu}\Delta_{2}(p;\mu)\right)\\
 & =-\frac{\mu^{-\epsilon}}{2}\frac{\lambda_{I}}{1-\lambda_{I}\frac{3}{2(4\pi)^{2}}\log\left(\frac{\mu}{\mu_{0}}\right)}\mu^{\epsilon}\int\frac{d^{d}p}{(2\pi)^{d}}\left[\frac{\epsilon}{p^{2}+\Gamma_{2}^{(0)}(\mu)}\right.\\
 & \left.-(2+\epsilon)\left(\frac{\mu^{2}}{(p^{2}+\Gamma_{2}^{(0)}(\mu))^{2}}-\frac{\mu^{2}}{p^{2}(p^{2}+\Gamma_{2}^{(0)}(\mu))}\right)\right]\\
 & =-\frac{\lambda_{I}}{1-\lambda_{I}\frac{3}{2(4\pi)^{2}}\log\left(\frac{\mu}{\mu_{0}}\right)}\left(\frac{\Gamma_{2}^{(0)}(\mu)^{(1-\epsilon/2)}}{(4\pi)^{2-\frac{\epsilon}{2}}}\left(1+(2+\epsilon)\frac{\mu^{2}}{\Gamma_{2}^{(0)}(\mu)^{1+\epsilon/2}}\right)+\mathcal{O}(\epsilon)\right)\\
 & \overset{\epsilon\to0}{=}\frac{\lambda_{I}}{1-\lambda_{I}\frac{3}{2(4\pi)^{2}}\log\left(\frac{\mu}{\mu_{0}}\right)}\frac{\Gamma_{2}^{(0)}(\mu)}{(4\pi)^{2}}\left(1+2\frac{\mu^{2}}{\Gamma_{2}^{(0)}(\mu)}\right)=\lambda\frac{\Gamma_{2}^{(0)}(\mu)}{(4\pi)^{2}}\left(1+2\frac{\mu^{2}}{\Gamma_{2}^{(0)}(\mu)}\right)
\end{align*}
the solution to this equation with the initial condition,
\[
\Gamma_{2}^{(0)}(\mu_{0})=-m_{I}^{2}
\]
 is,
\[
\Gamma_{2}^{(0)}(\mu)=\frac{4}{3}e^{\frac{64\pi^{2}}{3\lambda}}\text{\ensuremath{\mu_{0}^{2}}}E_{\frac{1}{3}}\left(\frac{64\pi^{2}}{3\lambda}-2\log\left(\frac{\mu}{\text{\ensuremath{\mu_{0}}}}\right)\right)-\frac{32\sqrt[3]{2}\pi^{4/3}e^{\frac{64\pi^{2}}{3\lambda}}\text{\ensuremath{\mu_{0}^{2}}}E_{\frac{1}{3}}\left(\frac{64\pi^{2}}{3\lambda}\right)+24\sqrt[3]{2}\pi^{4/3}m_{I}^{2}}{3\left(32\pi^{2}-3\lambda\log\left(\frac{\mu}{\text{\ensuremath{\mu_{0}}}}\right)\right)^{2/3}}
\]
where $E_{\frac{1}{3}}$ denotes the exponential integral function
with index $1/3.$ It is noted that the effective squared mass $m^{2}=-\Gamma_{2}^{(0)}(\mu)$
for fixed $\mu<\mu_{0}$, decreases as a function of $\lambda_{I}$%
. The effective coupling $\lambda$ and mass squared $\frac{m^{2}}{\mu_{0}^{2}}$
are plotted as a function of $\frac{\mu}{\mu_{0}}$ in the range $0.001$
to $0.3$,
\begin{figure}[H]
\begin{centering}
\includegraphics[scale=0.3]{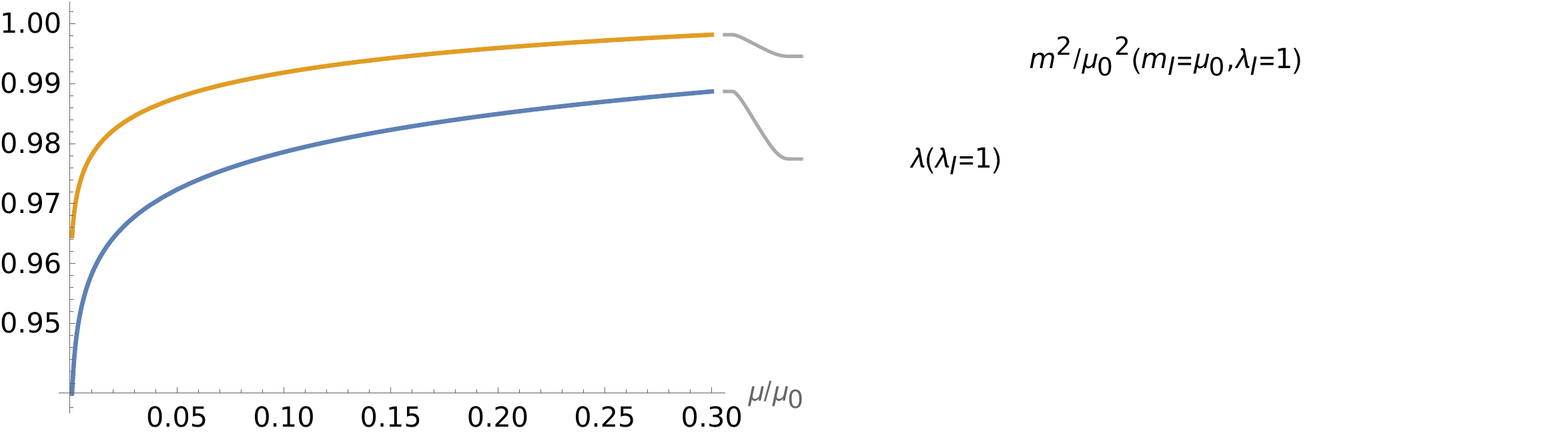}
\par\end{centering}
\caption{The effective coupling and mass squared for initial values $\lambda_{I}=1,\,m_{I}^{2}=\mu_{0}^{2}$,
are plotted as a function of $\frac{\mu}{\mu_{0}}$ in the range $0.001$
to 0.3}
\end{figure}

\noindent the above shows that these results are consistent with assuming
a low momenta dependence of these quantities.

Higher order terms in the low momentum expansion can be computed.
For example for the case of $\Gamma_{4}$ the momentum integrals that
appear are of the form,
\[
\int\frac{d^{d}p}{(2\pi)^{d}}\frac{p^{\mu_{1}}\cdots p^{\mu_{n}}}{(p^{2}+m_{I}^{2})((p+P)^{2}+m_{I}^{2})}
\]
these integrals can be computed in the usual way, i.e. using Feynman
trick,
\[
\frac{1}{AB}=\int_{0}^{1}\frac{dx}{[Ax+B(1-x)]^{2}}
\]
and the result for the momentum integrals of the form,
\[
F_{n}^{\mu_{1}\cdots\mu_{n}}=\int\frac{d^{d}p}{(2\pi)^{d}}\,\frac{p^{\mu_{1}}\cdots p^{\mu_{n}}}{(p^{2}+2p\cdot k+C)^{\alpha}}
\]
this calculation will not be done in this work. The above comments
are meant to show that their calculation, although lengthy, presents
no additional difficulties. 

\section{Conclusions and Outlook}

Conclusions and further research motivated by this work are summarized
in the series of remarks given below, 
\begin{itemize}
\item This work shows that dimensional regularization(DR) can be employed
in studying the WRG .
\item It is shown that DR can be though as a particular soft way of separating
high and low momenta degrees of freedom.
\item These results have been shown to hold without employing any approximation.
\item A particular relation between the scale parameter $\mu$ and the deviation
$\epsilon$ from dimension $4$ is required to obtain coincidence
of the solutions of the ERG equations with the field theory diagrammatic
contributions for the $k$ point proper function. This relation is,
\[
\frac{1}{\epsilon}=-\frac{S_{n-1}}{S_{d-1}}\frac{2}{k\gamma}\left(1-(\mu/\mu_{0})^{\frac{k}{2}\gamma}\right)
\]
although exact this relation is not universal. This relation shows
that the limits $\mu\to0$ and $\epsilon\to0^{-}$ are equivalent.
\item The application of this ERG formulation to gauge theories is a natural
and very interesting next step.
\end{itemize}

\section*{Appendix A}

\emph{The soft cut-off version of $\theta(p^{2}/\mu^{2}-1)$ leading
to to dimensional regularization. }
\begin{prop}
\ 

\[
\int_{0}^{\infty}dpf(p)\theta(\frac{p^{2}}{\mu^{2}}-1)=\int_{\mu}^{\infty}dp\,f(p)
\]
\end{prop}

\begin{proof}
\ 
\begin{align*}
\int_{0}^{\infty}dpf(p)\theta(\frac{p^{2}}{\mu^{2}}-1) & =\int_{0}^{\infty}\frac{d\left(p^{2}\right)}{2p}f(\sqrt{p^{2}})\theta(\frac{p^{2}}{\mu^{2}}-1)\\
 & =\int_{\mu^{2}}^{\infty}\frac{d\left(p^{2}\right)}{2\sqrt{p^{2}}}f(\sqrt{p^{2}})=\int_{\mu}^{\infty}dpf(p)
\end{align*}
\end{proof}
\begin{prop}
\label{prop:3}\ 
\[
\int_{0}^{\infty}dp\,f(p)\theta(\frac{p^{2}}{\mu^{2}}-1)=\lim_{\lambda\to0}\int_{0}^{\infty}dp'\frac{p'}{\sqrt{p'^{2}+\mu^{2}}}\left(\frac{p'^{2}}{\mu^{2}}+1\right)^{\lambda}f(\sqrt{p'^{2}+\mu^{2}})
\]
\end{prop}

\begin{proof}
\ 
\begin{align*}
\int_{0}^{\infty}dp\,f(p)\theta(\frac{p^{2}}{\mu^{2}}-1) & =\lim_{\lambda\to0}\int_{1}^{\infty}d(p/\mu)\mu\left(\frac{p^{2}}{\mu^{2}}\right)^{\lambda}f(p)\\
 & =\lim_{\lambda\to0}\int_{1}^{\infty}\frac{\mu^{2}}{2p}d(p^{2}/\mu^{2})\left(\frac{p^{2}}{\mu^{2}}\right)^{\lambda}f(\sqrt{p^{2}})
\end{align*}
in the last equality it was used that,
\[
d(p/\mu)=\frac{\mu}{2p}d(p^{2}/\mu^{2})
\]
making the change of variables,
\[
p'^{2}/\mu^{2}=p^{2}/\mu^{2}-1\Rightarrow p^{2}=p'^{2}+\mu^{2}
\]
then,
\begin{align*}
\lim_{\lambda\to0}\int_{1}^{\infty}\frac{\mu^{2}}{2p}d(p^{2}/\mu^{2})\left(\frac{p^{2}}{\mu^{2}}\right)^{\lambda}f(\sqrt{p^{2}}) & =\lim_{\lambda\to0}\int_{0}^{\infty}d(p'^{2}/\mu^{2})\frac{\mu^{2}}{2\sqrt{p'^{2}+\mu^{2}}}\left(\frac{p'^{2}}{\mu^{2}}+1\right)^{\lambda}f(\sqrt{p'^{2}+\mu^{2}})\\
 & =\lim_{\lambda\to0}\int_{0}^{\infty}dp'\frac{p'}{\sqrt{p'^{2}+\mu^{2}}}\left(\frac{p'^{2}}{\mu^{2}}+1\right)^{\lambda}f(\sqrt{p'^{2}+\mu^{2}})
\end{align*}
\end{proof}
\begin{prop}
\ 
\begin{align*}
\int d^{n}p\,\theta(\frac{p^{2}}{\mu^{2}}-1)f(p,q) & =\frac{S_{n-1}}{S_{d-1}}\int d^{d}p\,p^{\epsilon}\left(1+\frac{p^{2}}{\mu^{2}}\right)^{-\frac{\epsilon}{2}}\left(\left(1+\frac{\mu^{2}}{p^{2}}\right)^{\frac{d-1+\epsilon}{2}}\frac{p}{\sqrt{p^{2}+\mu^{2}}}f(\sqrt{p^{2}+\mu^{2}},q)\right)
\end{align*}
\end{prop}

\begin{proof}
\ 
\begin{align*}
\int d^{n}p\,\theta(\frac{p^{2}}{\mu^{2}}-1)f(p,q) & =\frac{S_{n-1}}{S_{d-1}}\int d^{d}p\,p^{\epsilon}\theta(\frac{p^{2}}{\mu^{2}}-1)f(p,q)\\
 & =\frac{S_{n-1}}{S_{d-1}}\int d\Omega_{n-1}\int_{0}^{\infty}dp\,p^{d-1+\epsilon}\theta(\frac{p^{2}}{\mu^{2}}-1)f(p,q)
\end{align*}
using \ref{prop:3} leads to,
\begin{align*}
\int d^{n}p\,\theta(\frac{p^{2}}{\mu^{2}}-1)f(p,q) & \overset{\lambda=-\epsilon/2}{=}\lim_{\epsilon\to0}\frac{S_{n-1}}{S_{d-1}}\int d\Omega_{n-1}\int_{0}^{\infty}dp\frac{p}{\sqrt{p^{2}+\mu^{2}}}\left(\frac{p^{2}}{\mu^{2}}+1\right)^{-\frac{\epsilon}{2}}(p^{2}+\mu^{2})^{\frac{d-1+\epsilon}{2}}f(\sqrt{p^{2}+\mu^{2}},q)\\
 & =\lim_{\epsilon\to0}\frac{S_{n-1}}{S_{d-1}}\int d\Omega_{n-1}\int_{0}^{\infty}dpp^{d-1+\epsilon}\frac{p}{\sqrt{p^{2}+\mu^{2}}}\left(\frac{p^{2}}{\mu^{2}}+1\right)^{-\frac{\epsilon}{2}}(1+\frac{\mu^{2}}{p^{2}})^{\frac{d-1+\epsilon}{2}}f(\sqrt{p^{2}+\mu^{2}},q)\\
 & =\lim_{\epsilon\to0}\frac{S_{n-1}}{S_{d-1}}\int d^{d}pp^{\epsilon}\left(\frac{p^{2}}{\mu^{2}}+1\right)^{-\frac{\epsilon}{2}}\left(\left(1+\frac{\mu^{2}}{p^{2}}\right)^{\frac{d-1+\epsilon}{2}}\frac{p}{\sqrt{p^{2}+\mu^{2}}}f(\sqrt{p^{2}+\mu^{2}},q)\right)
\end{align*}
\end{proof}
\begin{rem}
\ Using that,
\[
p^{\epsilon}\left(\frac{p^{2}}{\mu^{2}}+1\right)^{-\frac{\epsilon}{2}}=\mu^{\epsilon}\left(1+\frac{\mu^{2}}{p^{2}}\right)^{-\frac{\epsilon}{2}}=\mu^{\epsilon}\left(1+\frac{\epsilon}{2}\frac{\mu^{2}}{p^{2}}+\frac{\frac{\epsilon}{2}(\frac{\epsilon}{2}+1)}{2}\left(\frac{\mu^{2}}{p^{2}}\right)^{2}+\cdots\right)
\]
which converges for $\mu<p$, leads to,
\begin{align}
\int d^{n}p\,\theta(\frac{p^{2}}{\mu^{2}}-1)f(p,q) & =\lim_{\epsilon\to0}\frac{S_{n-1}}{S_{d-1}}\int d^{d}p\mu^{\epsilon}\left(f\left(p,q\right)+\frac{\mu^{2}}{2p^{2}}\left(pf^{(1,0)}\left(p,q\right)-2(\epsilon-1)f\left(p,q\right)\right)\right.\nonumber \\
 & \left.+\frac{\mu^{4}}{8p^{4}}\left(4(\epsilon-1)\epsilon f\left(p,q\right)+p(3-4\epsilon)f^{(1,0)}\left(p,q\right)+p^{2}f^{(2,0)}\left(p,q\right)\right)+\cdots\right)\label{eq:thetaf}
\end{align}
and,
\begin{align}
\int d^{n}p\,\left(\mu\frac{\partial}{\partial\mu}\theta(\frac{p^{2}}{\mu^{2}}-1)\right)f(p,q) & =\epsilon\int d^{n}p\,\theta(\frac{p^{2}}{\mu^{2}}-1)f(p,q)+\nonumber \\
 & +\lim_{\epsilon\to0}\frac{S_{n-1}}{S_{d-1}}\int d^{d}p\mu^{\epsilon}\left(\frac{\mu^{2}}{p^{2}}\left(pf^{(1,0)}\left(p,q\right)-2(\epsilon-1)f\left(p,q\right)\right)+\cdots\right)\label{eq:dthetaf}
\end{align}
\end{rem}

\section*{Appendix B}

\emph{The general solution to the ERG equation}. This equation is
given by,
\[
\left[\mu\frac{\partial}{\partial\mu}-p\cdot\frac{\partial}{\partial p}-k\frac{\gamma}{2}\right]\Gamma_{k}(\{p\},m;\mu)=L_{k}(\{p\},m;\mu)
\]
noting that,
\[
\sum_{i}p_{i}\cdot\frac{\partial}{\partial p_{i}}F(\{p\},m,\mu)=\left.\rho\frac{\partial}{\partial\rho}F(\rho\{p\},m,\mu)\right|_{\rho=1}
\]
a solution of the above equation is obtained from the solution to
the following equation,
\begin{equation}
\left[\mu\frac{\partial}{\partial\mu}-\rho\frac{\partial}{\partial\rho}-k\frac{\gamma}{2}\right]\Gamma_{k}(\rho\{p\},m;\mu)=L_{k}(\rho\{p\},m;\mu)\label{eq:eqrhomu}
\end{equation}
it is useful to change variables from $\rho,\mu$ to $r,\mu$ where
$r=\rho/\mu$, this is so because,
\[
-r\frac{\partial}{\partial r}=\mu\frac{\partial}{\partial\mu}-\rho\frac{\partial}{\partial\rho}
\]
in terms of the variables $r,\mu$ , the equation (\ref{eq:eqrhomu})
is,
\[
\left(-r\frac{\partial}{\partial r}-k\frac{\gamma}{2}\right)\Gamma_{k}(r\mu\{p\},m;\mu)=L_{k}(r\mu\{p\},m;\mu)
\]
a general solution to this equation is given by,
\[
\Gamma_{k}(r\mu\{p\},m;\mu)=r^{-k\frac{\gamma}{2}}\left(\Phi(\mu\{p\},m)-\int_{r_{0}}^{r}\frac{d\tilde{r}}{\tilde{r}^{1-k\frac{\gamma}{2}}}L_{k}(\tilde{r}\mu\{p\},m;\mu)\right)
\]
 taking $\rho=1$, the result in terms of $\mu$ is obtained replacing
in the previous equation $r\overset{\rho=1}{=}\mu^{-1}$. Leading
to,
\[
\Gamma_{k}(\{p\},m;\mu)=\mu^{k\frac{\gamma}{2}}\left(\Phi(\mu\{p\},m;\mu)-\left[\int_{r_{0}}^{r}\frac{d\tilde{r}}{\tilde{r}^{1-k\frac{\gamma}{2}}}L_{k}(\tilde{r}\mu\{p\},m;\mu)\right]_{r=\mu^{-1}}\right)
\]

\section*{Appendix C}

\emph{The expression,
\[
\Gamma_{k}(\{p\},m,\mu)=\mu^{\frac{k}{2}\gamma}\left(\Phi(\{p\}\mu,m\})-\tilde{\Gamma}_{k}^{RD}(\{p\},m,\mu)\right)
\]
where,
\begin{align}
\tilde{\Gamma}_{k}^{RD}(\{p\},m,\mu)= & \frac{\mu^{-\epsilon}}{2}\int\frac{d^{d}p}{(2\pi)^{d}}\frac{\epsilon}{p^{2}+m^{2}}\left[\sum_{l=1}^{\infty}d^{d}p_{1}\cdots d^{d}p_{l}\right.\nonumber \\
 & \times\sum_{m_{1}\cdots m_{l}}\bar{\Gamma}_{m_{1}+2}(p,-p_{1},p_{1}^{(1)},\cdots,p_{m_{1}}^{(1)})\frac{1}{p_{1}^{2}+m^{2}}\nonumber \\
 & \times\bar{\Gamma}_{m_{2}+2}(p_{1},-p_{2},p_{1}^{(2)},\cdots,p_{2}^{(2)})\frac{1}{p_{2}^{2}+m^{2}}\nonumber \\
 & \left.\times\cdots\bar{\Gamma}_{m_{l}+2}(p_{k-1},-p,p_{1}^{(k)},\cdots,p_{m_{k}}^{(k)})\right]\label{eq:Lmo=0000F1o-1}
\end{align}
 coincides with the one obtained in the dimensionally regularized
field theory calculation. }

In order to show this last assertion it is noted that the above expression
for $\tilde{\Gamma}_{k}^{RD}$ is a summation over all possible diagrams
that contribute to this function, these diagrams being constructed
as a $1$-loop graph with proper vertices connected by $\Delta_{2}^{ft}(p;\mu)$
lines. Below it is shown that all the field theoretic diagrammatic
contributions to any proper function are of this form. In order to
see this, the following is important,
\begin{prop}
Any proper graph with a non-vanishing number of internal lines is
a 1-loop graph with proper vertices.
\end{prop}

\begin{proof}
Take a proper graph $G$ with any number of external lines and a non-vanishing
number of internal lines. Now consider cutting one internal line $l$
in this graph. The resulting graph could be,

(i) \emph{A tree graph}. Then this implies that the original graph
was a 1-loop graph. Indeed, if the graph $G$ is characterized by
having $V$ proper vertices and $I$ internal lines then the number
of loops $L$ is given by $L=I-V+1$ , if removing a internal line
gives a tree graph then $0=(I-1)-V+1\Rightarrow I=V$ and thus the
original graph is a 1-loop graph.

(ii) \emph{A proper graph $P$ which is a sub-graph of $G$}. Then
$G$ is a 1-loop graph as shown in the figure,
\end{proof}
\begin{center}
\includegraphics[scale=0.2]{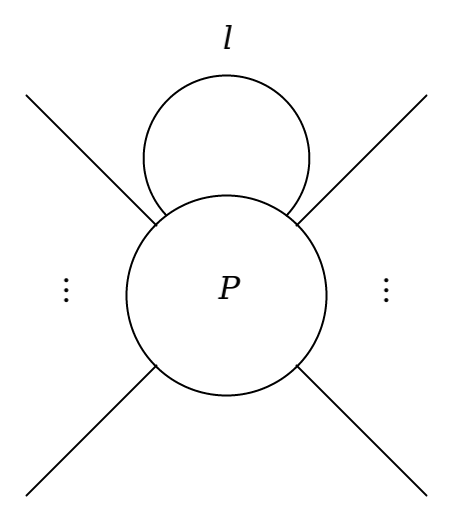}
\par\end{center}

Next it is shown that the proper vertices that appear in expressing
a given proper function $\Gamma_{k}$ as a $1$-loop diagram, are
exactly the ones involved in the summation appearing in the r.h.s.
of (\ref{eq:Lmo=0000F1o}). Let $v=1,\cdots,V$ denote the proper
vertices in $G$, and $l_{v}$ the number of lines emerging form the
proper vertex $v$. The the following relation holds,
\[
\sum_{v}l_{v}=k+2I\Rightarrow k=\sum_{v}(l_{v}-2)
\]
where $k$ denotes the number of external lines of the graph $G$.
The second equality follows from the fact that since, as shown above,
$G$ is a 1-loop graph, then $I=V$. Then this means that, 
\[
k=\sum_{v}m_{v}
\]
where $m_{v}+2=l_{v}$. Thus given $k$, the admissible values for
$m_{v}$ are just the ones involved in the summation appearing in
(\ref{eq:Lmo=0000F1o}). Then, the r.h.s. of (\ref{eq:Lmo=0000F1o})
consists in a summation over all possible proper vertices linked by
propagator $\mu^{-\epsilon}\Delta_{2}$ contributing to the the $k$-point
proper graph, this proves the assertion at the beginning section \ref{sec:Solutions-of-dimensionally}.

\vspace{0.5cm}

\begin{center}
\textbf{Acknowledgements. }
\par\end{center}

I am deeply indebted to G. Torroba for sharing his expertise on the
renormalization group, for many enlightening discussions and for a
critical reading of the manuscript. 

\bibliographystyle{unsrt}
\addcontentsline{toc}{section}{\refname}\bibliography{Bibliography,../../adsqcd/nf/Bibliography}

\begin{thebibliography}{1}

\bibitem{Wilson:1973jj}
K.~G. Wilson and John~B. Kogut.
\newblock {The Renormalization group and the epsilon expansion}.
\newblock {\em Phys. Rept.}, 12:75--200, 1974.

\bibitem{Wegner-PhysRevA.8.401}
Franz~J. Wegner and Anthony Houghton.
\newblock Renormalization group equation for critical phenomena.
\newblock {\em Phys. Rev. A}, 8:401--412, Jul 1973.

\bibitem{bagnuls:hal-00012738}
C.~Bagnuls and C.~Bervillier.
\newblock {Exact Renormalization Group Equations. An Introductory Review}.
\newblock {\em {Physics Reports}}, 348:91, 2001.
\newblock Final version. Many references added, section 4.2 added, minor
  corrections. 65 pages, 6 figs.

\bibitem{Weinberg1978}
Steven Weinberg.
\newblock {\em Critical Phenomena for Field Theorists}, pages 1--52.
\newblock Springer US, Boston, MA, 1978.

\bibitem{stuckelberg-petermann1953normalisation}
E.~C.~G. Stueckelberg and A.~Petermann.
\newblock La normalisation des constantes dans la th{\'e}orie des quanta.
\newblock {\em Helv. Phys. Acta}, 26:499--520, 1953.

\bibitem{Gellmann-low-PhysRev.95.1300}
M.~Gell-Mann and F.~E. Low.
\newblock Quantum electrodynamics at small distances.
\newblock {\em Phys. Rev.}, 95:1300--1312, Sep 1954.

\bibitem{bollini1972dimensional}
CG~Bollini and JJ~Giambiagi.
\newblock Dimensional renorinalization: The number of dimensions as a
  regularizing parameter.
\newblock {\em Il Nuovo Cimento B (1971-1996)}, 12(1):20--26, 1972.

\bibitem{tHooft:1972fi}
Gerard~'t Hooft and M.~J.~G. Veltman.
\newblock {Regularization and Renormalization of Gauge Fields}.
\newblock {\em Nucl. Phys.}, B44:189--213, 1972.

\bibitem{Morris-94}
Tim~R. {Morris}.
\newblock {The Exact Renormalization Group and Approximate Solutions}.
\newblock {\em International Journal of Modern Physics A}, 9(14):2411--2449,
  January 1994.

\end{thebibliography}

\end{document}